\documentclass{article}
\usepackage[utf8]{inputenc}
\usepackage{graphicx}
\usepackage[colorlinks]{hyperref}
\usepackage{pifont}
\usepackage{authblk}
\usepackage{longtable}
\usepackage{geometry}
\usepackage{lipsum}
\usepackage[utf8]{inputenc}
\usepackage{booktabs, makecell, longtable}

\title{A systematic literature review on Edge Computing security}
\graphicspath{Images/}
\begin{document}

\author[1]{Harsiddh Kalariya}
\author[2]{Vini Patel}
\author[3]{Kavish Shah}
\affil[1,2,3]{Engineering System and Computing/ Computer engineering, Graduate Student, University of Guelph}

\maketitle

\begin{table}[h!]
  \begin{center}
    \begin{tabular}{p{4cm} p{8cm}}
      \hline
      \textbf{ARTICLE INFO}  &  \textbf{ Abstract }\\
      \hline
      
      \textbf{2.}Edge-Computing Security
      
      \textbf{3.}Network Edge 
      
      \textbf{4.}Fog Computing 
      
      \textbf{5.}Edge-Computing Cyber-security 
      
      \textbf{6.}Edge Computing Cyber Security & Mobile and Internet of Things devices are generating enormous amounts of multi-modal data due to their exponential growth and accessibility. As a result, these data sources must be directly analysed in real time at the network edge rather than relying on the cloud. Significant processing power at the network's edge has made it possible to gather data and make decisions prior to data being sent to the cloud. Moreover, security problems have significantly towered as a result of the rapid expansion of mobile devices, Internet of Things (IoT) devices, and various network points. It's much harder than ever to guarantee the privacy of sensitive data, including customer information. This systematic literature review depicts the fact that new technologies are a great weapon to fight with the attack and threats to the edge computing security.\\
      \hline
    \end{tabular}
  \end{center}
\end{table}
\section{Introduction}

Edge Computing is a technology used to offload the tasks processed by the clouds. Hence, due to edge computing, clouds become more efficient in terms of bandwidth, privacy, operation cost, location awareness and so on \cite{wu2019edge, yazdinejad2020energy}. In simple words, edge computing enables us to process more data at the network edge. It could be near to the cloud services or near to the IOT or mobile device services\cite{shi2016promise,xiao2019edge,sha2020survey}. Due to the towering demands of IoT and mobile devices, the need of edge computing has become indispensable.  Hence, Edge computing is more suitable in most of the smart applications. This proliferated demand gives us more threats from the real-world \cite{xiao2019edge,a2,a3}. Some of the most common examples includes attacks caused by Malware\cite{abawajy2018identifying}, DDoS and Authentication attack \cite{xiao2019edge, myneni2022smartdefense,ren2018edge}.

However, before going in-depth about the security of the edge networks, let's get clear on why we need to have edge computing. Consider you are using an Instagram and you want to post a video. However, if you upload a raw video without being processed, it will occupy much data-space at the cloud network. In addition to this, it requires a lot of bandwidth and time to get this user-data to the cloud service. Myriad of users are  using Instagram and generating a lot of data. Now, if we process, compress and manage these data on user's phone, the size of the data can reduce significantly. This, in turn will improve the bandwidth management and cost operation. \cite{shi2016promise}. Similarly, Apple's Face-ID, Google's voice assistance are some of the examples of Edge Computing  which enhances security and user data privacy to a great extent. \cite{xiao2019edge, ranaweera2021survey,a5,satyanarayanan2017emergence, abbas2017mobile}. Also, Apple asks their developers to follow some rules to develop an application, which are again tested by the company itself. Later, these applications are allowed to be distributed. This is another great example of edge computing \cite{abawajy2018identifying,a4}.

There are confusing terms when discussing about the edge computing which are cloud computing, Cloud-lets and fog computing. When it comes to cloud, it is consider as the ultimate resource to provide services which manages all of the app data and are used to compute most of the complicated computational tasks. On the other hand, Cloud-lets are nothing but the clouds having less powers as compared to cloud and located near to the users' locations. Lastly, fog computing can be defined as the layer between the edge computing layer and the cloud or Cloud-lets computing layer \cite{dolui2017comparison, yousefpour2019all}. Sometimes, Fog computing and Edge computing are interchanged \cite{paharia2020comprehensive,shirazi2017extended}. Even though the term Edge-computing is confusing in its own, it can simply define as a computing methodology to improve network performance and cloud performance \cite{paharia2020comprehensive,caprolu2019edge}. 

By referring the major research papers, we can easily say that the need of edge computing will escalate in distinct future due to soaring use of Internet of Things (IoT) devices \cite{ ranaweera2021survey,satyanarayanan2017emergence,li2016mobile}. Thus, we may see copious cyber security attacks to these devices leading interruption, compromises or reduced quality services overall. This research paper will mainly focus on the most common prevention for the security attacks and latest methods and application by reviewing existing papers based on edge computing security attacks and their prevention. The ultimate goal is to provide better consequence driven by research community to give new insights on this topic.

\subsection{Prior Research}

When finding the work that has been performed on Edge computing security, we can say from our best knowledge, that there are not much research papers available due to its recent emergence. M. Satyanarayana has denoted that, the word edge computing started getting attention with the use of Content Delivery Networks (CDNs) in late 1990 \cite{satyanarayanan2017emergence}. Henceforth, we have seen escalated call in IoT devices. One of the another best research paper published by Yinhao Xiao and team \cite{wu2019edge}, stated different security attacks and their mechanism usually seen on edge computing IoT devices including DDoS attack, Side Channel attack, Malware Injuction attack and authentication attack. Similarly, Daojing He, Sammy Chan and Mohsen Guizani published a paper on security in IOT edge computing\cite{he2018security,ren2018edge,a6} wherein attacks are divided into passive attacks including \textbf{1.} eavesdropping and \textbf{2.}Traffic Analysis and active attacks including \textbf{1.} tampering with data packets, \textbf{2.} forgery packets, \textbf{3.} replay attacks and {4.} denial-of-services (DoS) attack.
    
According to an overview written on Edge computing research \cite{cao2020overview,alwarafy2020survey,a6}, data security and privacy protection has been one of the concurrent research content on edge computing. Pasika Ranaweera, Anca Delia and Madhusanka Liyanage stated one of the possible reason for the security threat to the edge computing. Due to high demand in IOT devices and to make IOT devices affordable, companies are making cheap circuitry which again give a rise to weak cryptography algorithms. The vulnerability of these IOT devices are again high to cloning, physical tampering and security issues of the devices \cite{ranaweera2021survey}. One article from ScienceDirect focused on the threats caused by malware. A malware code, which is written knowingly or unknowingly, can causes and leads to many serious issues in edge computing including but not limited to, \textbf{1.} security controls, \textbf{2.} privacy expectations, \textbf{3.} Denial of Service (DoS), \textbf{4.} spam/advertising etc \cite{abawajy2018identifying,potrino2019distributed,liao2019security}.

To provide a solution to the threats and security issues of edge computing, Tian Wang and team \cite{wang2020intelligent} has focused on the use of machine learning algorithms to solve the problems. We can use different machine learning algorithms including SVMs, K-NN, NN etc to solve the security issues in edge computing. \cite{singh2021machine,a7,wang2020convergence} Although, there are not much articles found in this category as both the machine learning and edge computing are newly emerged branches and hence required to have proper research and development phase before citing anything about it.

Multifarious studies have been performed to give the ideas on the threats and security issues that already exist in the case of edge computing. However, we know that, edge computing term is confusing in its definition. Thus, We can not expect to have proper solution for each problems. Here, we have tried to give an overview of the work that has been performed on the security issues and threats faced on the edge computing. 

\subsection{Research Goals}
The ultimate goal of this SLR is to research on existing papers focusing on common problems for edge computing and feasible solutions for such threats. To make things simple, we came up with a table listing three research questions as shown in \hyperref[tab:table1]{Table \ref*{tab:table1}}.

\begin{table}[ht!]
\caption{Research Questions}
\label{tab:table1}
  \begin{center}
    \begin{tabular}{p{5.5cm}    p{5.5cm}}
      \hline
      \textbf{ Research Questions (RQ)}  &  \textbf{ Discussion }\\
      \hline
      \textbf{RQ(1)}  What are the latest applications focused on edge computing security? & The latest applications of edge computing security are discussed. Also whether machine learning is useful or not for solving any problems related to edge computing security are reviewed and discussed. \\
      \\
      \textbf{RQ(2)} What are the most common types of attacks and what associated mitigating solutions targeting edge computing networks? & The most common types of attacks and their possible solutions using new different technologies are listed and discussed.\\
      \\
      \textbf{RQ(3)} What prospects exist for edge computing security and privacy researchers in terms of future research?  &  The need for "The proper definition of edge computing" is discussed. In addition, all new attacks require to be identified properly to provide, improve and manage edge computing effectively.  \\
      \\
      
      \hline
    \end{tabular}
  \end{center}
\end{table}

\subsection{Contributions and layout}
This SLR is interrelated with the research that is already being carried out and has the contributions towards the society as follow: 

\begin{itemize}
    \item We came up with primary 38 research papers and documents related with edge computing security listed up to mid 2022. Other researchers may refer this SLR to extend their works.
    \item Among selected 38 research papers, we filtered 27 research papers that are best fit for this SLR. These research papers can be used as a guidance to refer any of the inference.
    \item We used these 27 research papers and documents to conduct thorough review and based on the data presented, we came up with the ideas, inferences and conclusions on the topic of edge computing security.
    \item We mainly focused on the security threats to the edge computing and provided possible solutions available to mitigate such issues.
    \item By following these documents, we create a standard to provide support in any research work related with Edge Computing Security.
\end{itemize}

The format of this essay is as follows: The procedures used to choose the primary studies for analysis in a precise way are described in \hyperref[sec:ResearchMethodology]{Section 2}. The outcomes of all the primary research chosen are summarized in \hyperref[sec:Findings]{Section 3}. The findings in relation to the relatively early study topics are explored in \hyperref[sec:discussion]{Section 4}. The research is accomplished in \hyperref[sec:FutureWork]{Section 5}, which also makes some proposals for additional study.

\section{Research Methodology}
\label{sec:ResearchMethodology}
To fulfil the objectives of the research question listed above on edge computing, we followed a benchmark \cite{kitchenham2007guidelines} provided by Kitchenham and Charters to conduct this SLR. We have accompanied the planning phase, finding phase and conclusion phase to carried out comprehensive SLR on edge computing. 

\subsection{Selection of primary studies}
The initial phase of collecting documents and research papers started with the keywords passed to the searching services. To perform such operations, the Boolean operators including AND and OR are used. The sentence is formed in such a way to get the finest practicable inferences to support the answers of the questions listed in \hyperref[tab:table1]{Table \ref*{tab:table1}}. Hence the sentences used to feed search facilities were:
\begin{itemize}
    \item ("Edge Computing" OR "Edge-Computing") AND "Security"
    \item ("Edge Computing" OR "Edge-Computing" OR "Network Edge" OR "Fog Computing") AND ("Security" OR "Cyber-Security" OR "Cyber Security" OR "CyberSecurity")
\end{itemize}
The searched platform were:
\begin{itemize}
    \item[\ding{217}] IEEE Xplore Digital Library
    \item[\ding{217}] Google Scholar
    \item[\ding{217}] SpringerLink
    \item[\ding{217}] ScienceDirect
    \item[\ding{217}] ACM Digital Library
\end{itemize}

Depending on different search facilities, the strings are taken from the title of SLR, keywords as listed above or even from the abstract. All search queries are done on the 14th October 2022 and hence the related documents and research paper till date are processed and reviewed. To filtered out the resultant records, Inclusion and exclusion criteria are used to produce better fit to the SLR. Once the results are processed through  \hyperref[sec:IncExcCr]{Inclusion and exclusion criteria}, forward and backward snowballing process described by Wohlin \cite{wohlin2014guidelines} is applied to these findings. This process is an iterative process and it goes on till the \hyperref[sec:IncExcCr]{Inclusion criteria} fits to the research papers.

\subsection{Inclusion and exclusion criteria}
\label{sec:IncExcCr}
The first important criteria for the research paper to be considered for SLR is to have factual arguments and inferences which can be case studies, any new computing method application or any security article listing any new threats or feasible solutions to the vulnerability of the edge computing. More importance is given to the paper containing common security threats and their solutions. Secondly, the paper or document must be peer-reviewed and must be written in English. All documents/papers should display proper references that must be accessible till date. If it seems to be of lower-grade, it will be discarded. In addition, a document seems to be identical or having multiple versions, then the latest version will be considered. \hyperref[tab:table2]{Table \ref*{tab:table2}} shows all inclusion and exclusion criteria.

\subsection{Selection results}
While searching for the papers, 247 related studies are selected using the initial keywords. We removed 46 studies after running   \hyperref[sec:IncExcCr]{criteria} in which documents having duplicate or similar studies and some of them are not having proper references or references are not accessible. Furthermore, at the final stage, we came up with 38 research papers. These all 38 papers are read in details and on running \hyperref[sec:IncExcCr]{criteria} we are left with 23 papers for this SLR. At the end, we run snowballing and 3 more documents are added to the list having a final number of 27 in total.

\begin{table}[h!]
\caption{Inclusion and exclusion criteria for the primary studies.}
\label{tab:table2}
  \begin{center}
    \begin{tabular}{p{5.5 cm}    p{5.5 cm}}
      \hline
      \textbf{Criteria for Inclusion}  &  \textbf{Criteria for   Exclusion}\\
       \hline
        The findings must have factual inferences related to the applications on Edge Computing Security & The papers having legal impact or business documents or economic application on edge computing or edge computing security.\\
        The findings having recent or common cyber-attacks placed in edge computing security and their possible solutions.  &  Findings which are grey literature including government documents, blogs etc.\\
        \\The findings must be peer-reviewed articles and must be published in a conference proceeding and/or in journal. & If findings written in a language other than English.\\
        \hline
    \end{tabular}
  \end{center}
\end{table}

\subsection{Quality assessment}
The evaluation of the quality of the documents and primary findings were made using the guidelines given by Charters and Kitchenham\cite{kitchenham2007guidelines}. 

\begin{description}
    \item[Stage 1:] \textbf{Edge Computing}. The paper must have definition of edge computing. The document should focused on application or security of edge computing.
    \item[Stage 2:] \textbf{Context}. The document have enough context for making conclusions and inferences. It helps us to make accurate and precise decisions.
    \item[Stage 3:] \textbf{Edge Computing Security}. The documents must list any possible threats or vulnerability for network edge or edge computing. At this point, document may or may not have provided proper solution in its findings.
    \item[Stage 4:] \textbf{Edge Computing Security Solution}. The documents must list possible vulnerability plus proper and practical solution to that vulnerability or threat.
    \item[Stage 5:] \textbf{Edge Computing Solution Performance}. Compare different solutions for similar problems. It gives us a brief idea about the comparatively best solution.
    \item[Stage 6:] \textbf{Data acquisition}. The details about data acquired, processed and described must be mentioned. It helps us to get the idea on accuracy.
\end{description}

\begin{table}[h!]
\caption{Excluded Studies}
\label{tab:table3}
  \begin{center}
    \begin{tabular}{l@{}@{\hspace{2em}}@{\hspace{2em}}l}
      \hline
      \textbf{Criteria Checklist stages}  &  \textbf{Excluded Studies}\\
       \hline
        Stage 1: Edge Computing & [S27]\\
        Stage 2: Context & [S26]\\
        Stage 3: Edge Computing Security & [S30]\\
        Stage 4: Edge Computing Security Solution & [S29]\\
        Stage 5: Edge Computing Solution Performance & [S18]\\
        \hline
    \end{tabular}
  \end{center}
\end{table}

The steps listed above are the quality assessment checklist which will be applied to all findings.

\subsection{Data Extraction}
The data extraction must be applied to all finding that has passed the quality assessment test. However, to check whether the procedure of data extraction is suitable or not, we had applied this process to initial five findings. Once, we get the desire result, the data extraction process is applied to all articles. The data from each findings are collected, processed and stored in a spreadsheet.

\textbf{Context Data:} The data about the aim and ultimate goal of the study.

\textbf{Qualitative Data:} The documents and inferences provided by authors and reviewed by peer.

\textbf{Quantitative Data:} The data collected based on any statistics or any experimental studies.

\begin{figure}[ht]
  \centering
  \includegraphics[width=0.75\textwidth]{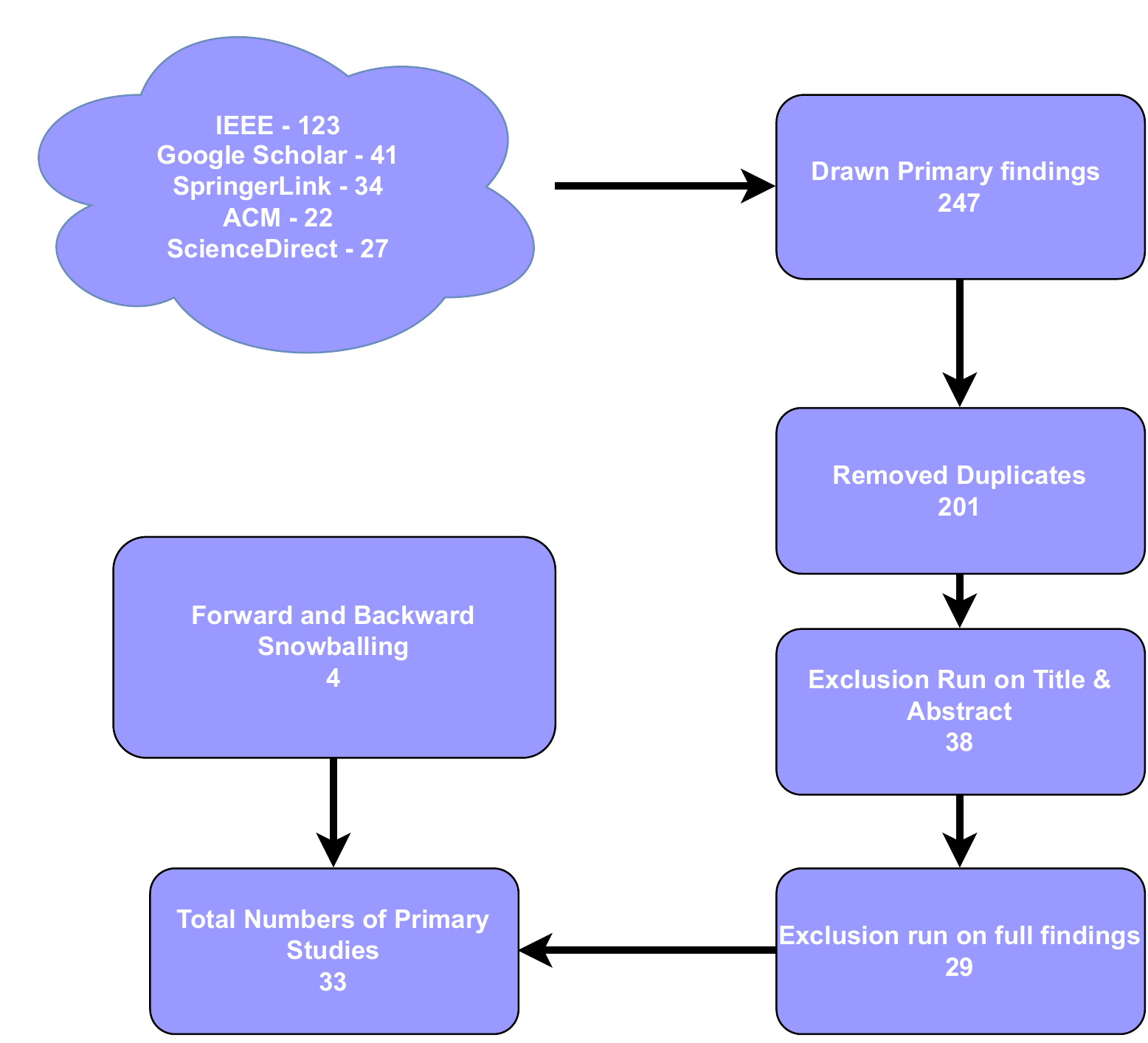}
  \caption{Collecting findings through processing}
  \label{fig:fig1}
\end{figure}

The detailed process on how findings are collected, processed and  referred stage by stage is shown in figure \ref{fig:fig1}

\subsection{Data Analysis}
To fulfil the objectives of discovering the answers of questions listed in \hyperref[tab:table1]{Table \ref*{tab:table1}}, we have put quantitative and qualitative data together. 

\subsubsection{Publications over time}
\begin{figure}[ht]
  \centering
  \includegraphics[width=0.75\textwidth]{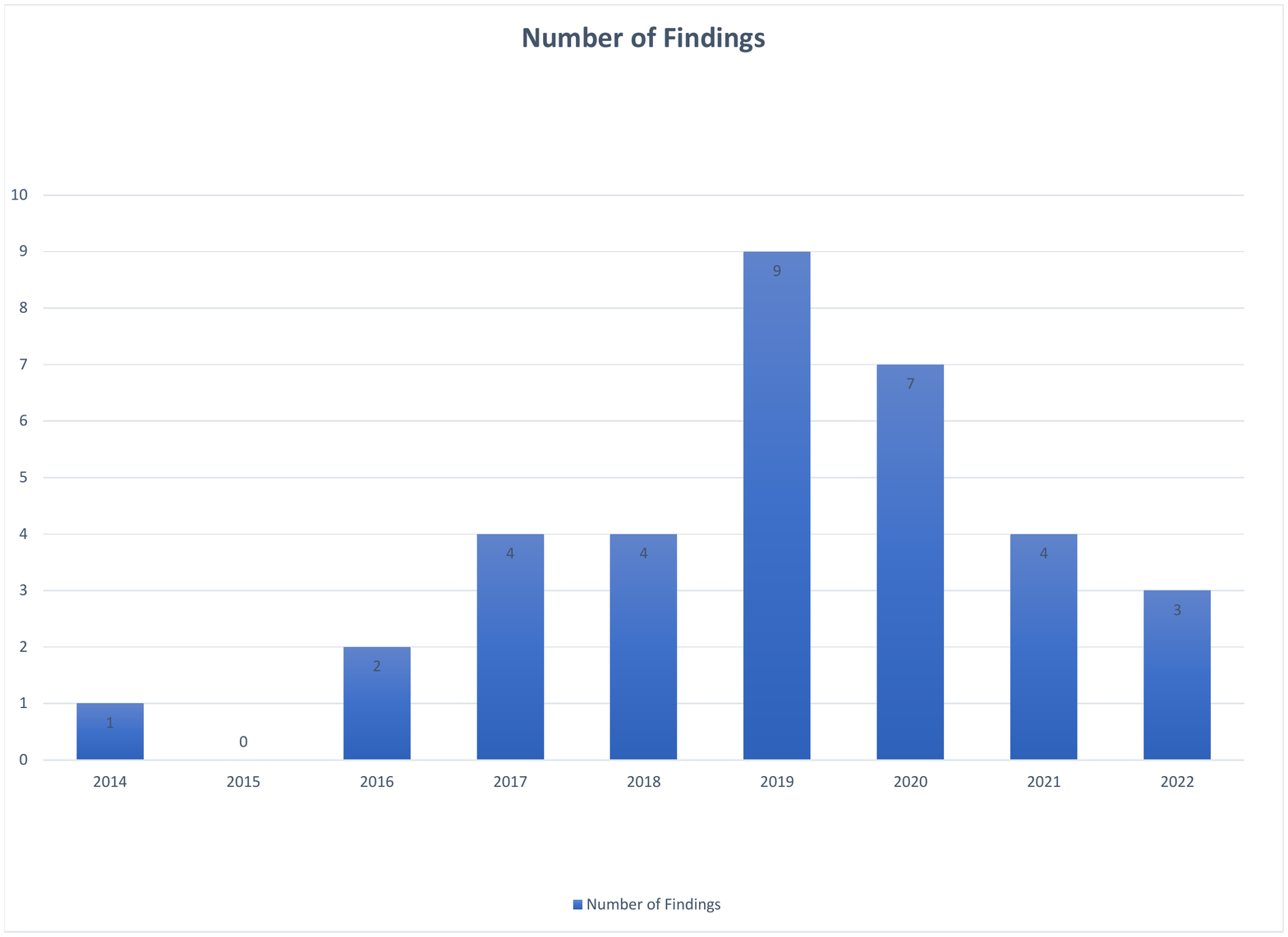}
  \vspace{-1em}
  \caption{{\footnotesize \textit{Collecting findings through processing}}}
  \label{fig:fig2}
\end{figure}

As we have stipulated the fact that, the branch and concept of edge computing is newly born and there is a lot of phenomenon researchers have to research on. The same fact can be referred from figure \ref{fig:fig2} which depicts the newness of edge computing especially when it comes to possible threats and security related to it. We can clearly see that after 2016, we have escalated numbers of related documents in the field of edge computing security. As the growth of IOT devices increase, the number of possible documents and findings related to the field will also increase giving more possible threats and security attacks and their prevention using latest technologies.

\subsubsection{Significant Keyword Count}
In order to find a common pattern among the foremost findings, we run an information retrieval test to list the statistics among the documents. The resultant phenomena is listed as a part of the \hyperref[tab:table4]{Table \ref*{tab:table4}}. By referring it, we can conclude that, Data and Protection are the most appearing words after excluding the title word Edge and Security. The word "Computing" on the other hand listed on the 9th rank, which we expected to be in the top five list.

\begin{table}[h!]
\caption{Keywords frequencies in the Primary Studies}
\label{tab:table4}
  \begin{center}
    \begin{tabular}{l@{}@{\hspace{2em}}@{\hspace{2em}}@{\hspace{2em}}@{\hspace{2em}}l}
      \hline
      \textbf{Keywords}  &  \textbf{Count}\\
       \hline
        Edge & 1787\\
        Data & 1568\\
        Security & 1356\\
        Protection & 1007\\
        cloud & 966\\
        Attack & 855\\
        Devices & 831\\
        Network & 775\\
        Computing & 724\\
        Authentication & 699\\
        Defense & 667\\
        Devices & 642\\
        Mobile & 559 \\
        Detection & 481\\
        Permission & 379\\
        Privacy & 274\\
        Protocol & 243\\
        Information & 153\\
        \hline
    \end{tabular}
  \end{center}
\end{table}

\clearpage

\section{Findings}
\label{sec:Findings}
\begin{figure}[ht]
  \centering
  \includegraphics[width=0.75\textwidth]{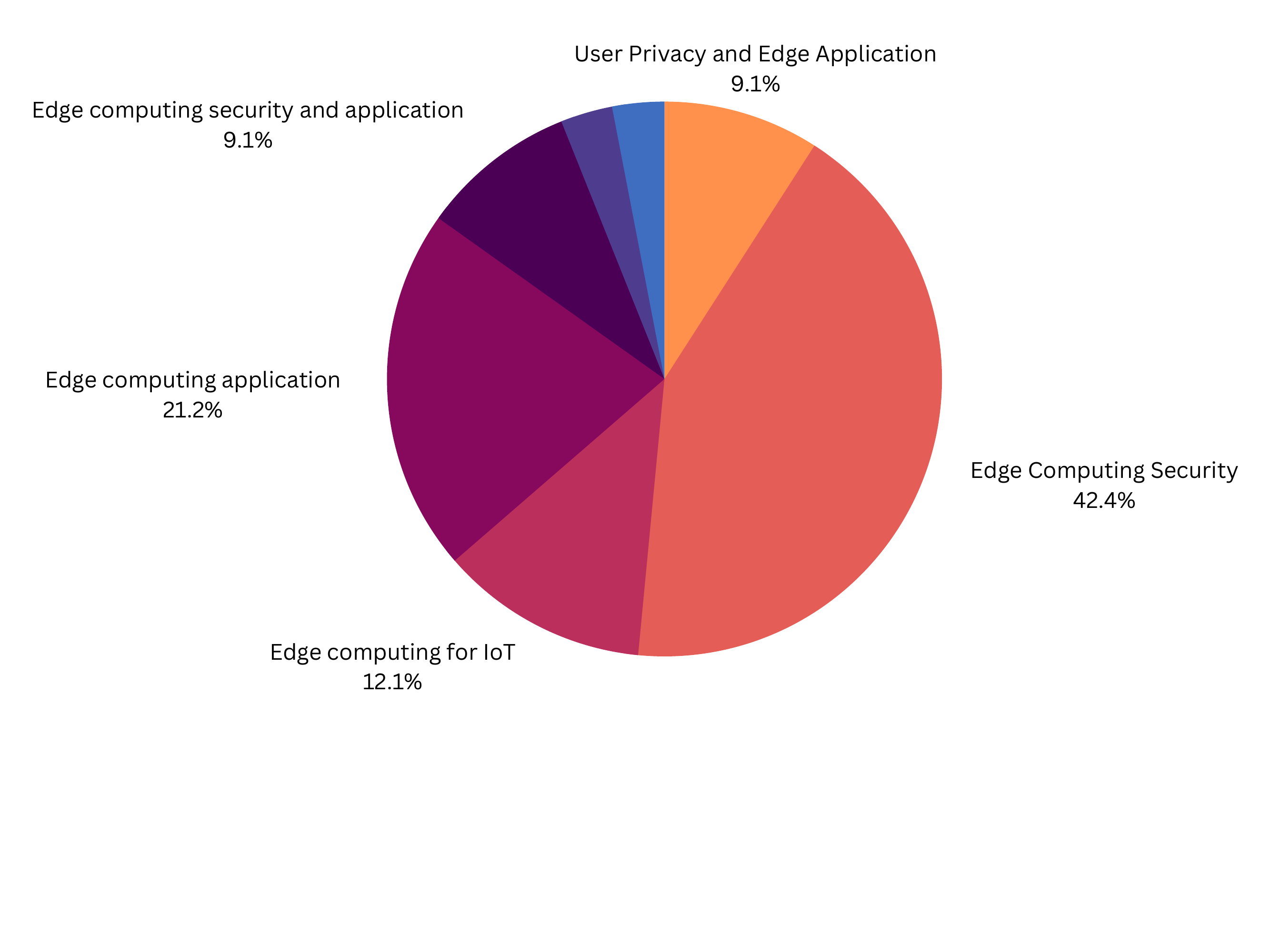}
  \vspace{-1em}
  \caption{{\footnotesize \textit{Chart representing primary studies}}}
  \label{fig:fig3}
\end{figure}

All relevant findings and the themes of each of the paper has been studied and the qualitative and quantitative data about the papers has been listed in figure \ref{fig:fig3}.
These studies are done keeping in mind the common attacks and problems occur in Edge Computing and possible solutions that can be given using machine learning or any new emerging technologies.

Furthermore, to simplify the study, each of the document were classified in different categories. As an example, primary study focusing of Edge Computing security are classified into \textbf{security} category. Similarly, study focusing on edge computing application is classified under the category of \textbf{Edge application}. Moreover, the paper is focusing on finding solution of different attacks on edge computing using different solution. Hence, we can clearly see that, almost 43\% documents are focused on edge computing security. Additionally, we got 21\% and 12\% for \textbf{Edge computing application} and \textbf{Edge computing for IoT}. 

The statistics of this findings is illustrated in the \hyperref[tab:table5]{Table \ref*{tab:table5}}. The finding includes study on security of edge computing, what are the attacks and threats causing security issues to the edge computing and any feasible findings that can be useful to improve the securities. Some studies are focused on the applications of the edge computing listing why edge computing is needed, what are the problems can be solved using edge computing, how network security will be improved with the help of edge computing and so on. The problem on the other hand, can be handled with better way if latest and emerging technology such as machine learning can be used to solve the problem. Most common attacks such as network outage attack, DDoS attack, user privacy breach attacks, malware injection attacks, side channel attack can be tackle down with the help of AI and/or ML.

\section{Discussion}
\label{sec:discussion}
Initially, while searching with the security of edge computing, the number of findings were retrieved focusing on what edge computing is, why edge computing is needed, what are the problems edge computing can be solved. As a result, we found out the fact, that the edge computing is the architecture used to bring cloud architecture more near to the user end. Hence, edge computing can be done on any side, either on cloud end or at user end. However, we came across many cyber-security threats causing a damage to either edge computing individually or as a whole communication link of edge computing. Edge computing suffers from the common attacks including but not limited to; DDoS attack, side channel attack, malware injection attack, Man-in-the-middle attack, authentication and/or authorization attack and so on. Machine learning on one hand or on other hand can improve the Quality of Service(QoS), stability, privacy, latency and so on.

For an example, to detect attacks caused by malicious behaviours can be detected using supervised machine learning algorithm. In this technique, malicious data with appropriate labels can be used to train algorithm to monitor and detect malicious data. Once, our algorithm is trained properly, it can easily identify such data \cite{sedjelmaci2021trusted,a8,a9}. Scientists have also found the use of unsupervised machine learning to extract features and pattern from complex edge data. Thus, we can get the pattern of attack or even the cause of attack with the help of unsupervised machine learning\cite{sun2019ai}. In addition, machine learning algorithm can be used for offloading in edge computing with the help of reinforcement learning. Given 'm' number of devices that needed to offload a task with 'n' number of servers with different accuracy, reinforcement learning algorithm can be trained to maximize the output in terms of cumulative rewards\cite{sun2019ai}.

Machine learning algorithms are used to even mitigate the effect of DDoS attack using CNN. Some data are extracted and prepossessed to train neural network. This neural network, thus, able to detect famous Mirai botnet which was one of the cause for DDoS attack\cite{hwang2019detecting}. Also, different machine learning algorithms can be used to detect and identify spoofing attack such as KNN, SVM, decision tree, Naive Bayes and so on\cite{singh2021machine,a10}. By referring primary studies, we can put it in a nutshell that even being a new technology, machine learning can be used widely to solve age-old problems with a different perspective. Although, the lack of the clear document is one of the most common issue in using new technologies, with increased research and development, we will have optimized solution for such edge computing problems to make communication link stable.
 
\subsection{RQ(1)  What are the latest applications focused on edge computing security?}
Though edge computing is a broad term and it is still applicable at both of the end, i.e. User End (UE) and Cloud End (CE), edge computing application is applicable at both of these ends. As the need of edge computing is increased due to increased demand of IoT devices in 21st century, most optimized solutions are needed to make the connection stable and efficient in terms of quality of service(QoS). Also, the use of machine learning can be so useful in solving problems seen in the edge security. Researchers have used supervised learning, unsupervised learning and reinforcement learning to solve and identify different security issues, attacks and optimal ways to make edge computing more reliable.  

The latest studies found the application of edge computing security as follow:
\begin{itemize}
    \item IoT Device optimization --- Need to optimize applications and user data, so that it can use less bandwidth, data security and increase efficiency. Need to add edge computing in terms of fog computing or cloudlets which can help to reduce latency and help to reduce offloading of the task.
    \item Edge Computing Network Security --- Need of adding more verification layer using convolution methods or using new technology such as machine learning. It makes network robust against common cyber-security attacks on the network edge.
    \item Edge security on cloud side --- Use different machine learning algorithm such as reinforcement learning  to make cloud computing optimized. It includes increase in cloud efficiency in terms of task offloading, cloud security in terms of different malware attacks, cloud management in terms of bandwidth and different cloud resources.
    \item Edge computing security trend --- Identify new different patterns of attacks including the type of attack, the origin of attack if possible, the possible solution to the identified trends and attacks which includes use of unsupervised machine learning to extract information.
    \item User Data Privacy --- Use more decentralized structure to make user data more private. It includes use of different algorithms and structure to store user data locally, use of more secure and efficient data storing method to improve latency involved in edge computing.
\end{itemize}

\subsection{RQ(2) What are the most common types of attacks and what associated mitigating solutions targeting edge computing networks?}
Edge computing is considered as a way to offload tasks from cloud and bring cloud closer to the user end. Normally, edge computing has been proven efficient on its expectation. However, due to attacks caused on either network side, user end side or on cloud end side, the performance of the edge computing is affected directly or indirectly. To mitigate the effect of attacks, it is more important to identify such common attacks. Not only identifying the patterns and problems of the attack, but the origin or cause of the problem is equally important to mitigate the solution. This can be again any of the cause such as vulnerable code, security issue, architecture of the network or any other reason\cite{xiao2019edge}. The most common types of attacks and the solutions to such problems are listed below.
\begin{itemize}
    \item DDoS --- Distributed denial of service attack is one of the most common type of attack seen when it comes to edge computing. One of the idea is to use \textit{SmartDefense} method in which one DDoS defense module is being used on both customer side (known as C-defense module) and ISP side (known as P-defense module). This module will be programmable and will help ISP to know and mitigate the effect of DDoS attack\cite{myneni2022smartdefense}. C-defense can be implemented with the help of DNN which can be trained regularly to mitigate DDoS.
    \item Malware injection Attacks --- Network devices such as IoT devices, mobile devices or cloud servers can be injected with malware, which is nothing but the malicious software that harms security, efficiency or intigrity of the system. The possible cause of this attack can be found by running software and measuring its performance. If the performance of the device seems to be compromised, the factors causing the effect can be listed and the cause can be identified either manually or using unsupervised machine learning algorithm\cite{abawajy2018identifying,myneni2022smartdefense}.
    \item Authorization and/or Authentication attack --- These kinds of attacks seem more common in IoT devices as such devices may have weak credentials as a security parameters. In addition, IoT devices have many ways to transfer information such as Bluetooth, Wi-Fi, other wired protocol such as USB or any other related sensors. If any of these interface is compromised, it can lead to such kind of attack\cite{xiao2019edge}. One simple solution is to use strong credentials for authentication or use more secure method for authentication such as fingerprint sensor or face detection. Changing password or other credentials regularly seems simple but effective solution to such problem.
    \item Man-in-the-middle(MITM) attack --- Though MITM attack usually seen as a cyber-security attack, it is one of the common attack occurs on the edge computing security as well\cite{xiao2019edge}. MITM can be solved using supervised machine learning algorithm as it will be easy to collect such data and label by performing attack artificially. One finding shows similar action by performing supervised machine learning to detect MITM attack\cite{al2020man}.
    \item Other Security risks --- Apart from the attacks listed above, various attacks are seen causing damage to the edge computing such as data injection attack, which usually leads to poor system performance, side-channel attack, which usually involves hardware parameters to debug, analyse and make some conclusion to draw some information from the system and so on. There are some findings describing the solutions, but more or less, information that satisfy proper solutions to such problems have not been retrieved and hence can be work upon in future. 
\end{itemize}

\subsection{RQ(3) What prospects exist for edge computing security and privacy researchers in terms of future research}
Security of the networking system is a challenging task whether it is cloud computing security, edge computing security or any related system security. In terms of edge computing however, we know that there are many different kinds of attacks which causes serious damage to our systems. While talking about the prevention of the attacks many approaches come into picture including age old method to solve problem to new solution of using machine learning. Although, old methods work, but machine learning provides optimal feasible solution to the problem. Being a new branch, we are still seeking as much resources as possible \cite{myneni2022smartdefense,unal2022machine,chen2019deep,hou2019use,sedjelmaci2021trusted,sun2019ai,hwang2019detecting,al2020man}, however those are not sufficient. Thus, as time passes, we will be having more number of research and experimental results to such problems using machine learning.

We have seen most types of machine learning algorithms in action including supervised machine learning, unsupervised machine learning and reinforcement learning for providing needed security and privacy to the user end and to the cloud end \cite{sun2019ai}. In future, we will be watching how well these algorithms are performing. The growth of the machine learning is again connected with the growth in the data. Hence, time will generate the required data for the experiment! 

\begin{longtable}{{p{2 cm} p{6 cm} p{4 cm}}}
    \caption{Findings and themes of the primary studies.}
    \label{comp}        \\
    \toprule
    Primary Study &   \makecell[b]{Qualitative and Quantitative \\ Key Data Reported} &   \makecell[b]{Application Category}  \\
    \midrule
\endfirsthead
    \caption[]{Findings and themes of the primary studies.}    \\
    \toprule
        Primary Study &   \makecell[b]{Qualitative and Quantitative \\ Key Data Reported} &   \makecell[b]{Application Category}  \\
    \midrule
\endhead
    \midrule
    \multicolumn{3}{r}{\footnotesize\itshape Continue on the next page}
\endfoot
    \bottomrule
\endlastfoot
\label{tab:table5}
        \hyperlink{[S1]}{[S1]} & How edge computing can be used to improve the privacy of the user as well as to reduce the latency and save bandwidth of the system & User Privacy and Edge Application\\
        
        \hyperlink{[S2]}{[S2]} & Different attacks on edge computing security, their types and their root causes. & Edge Computing Security \\

        \hyperlink{[S3]}{[S3]} & Identifying different malware and threats to the IoT devices which indirectly affects security of Edge Computing. & Edge computing for IoT\\

        \hyperlink{[S4]}{[S4]} & How IoT and edge computing sometimes refer together and what are the possible threats for IoT devices. & Edge computing for IoT\\

        \hyperlink{[S5]}{[S5]} & How edge computing is useful to increase efficiency of cloud by reducing latency, to improve security of clouds. Other applications of edge computing is also discussed. & Edge Application\\

        \hyperlink{[S6]}{[S6]} & How edge computing is used to reduce ingress bandwidth, to mask the cloud or network failures,to lower the management cost of centralized infrastructure.& Edge computing application 
        \\

        \hyperlink{[S7]}{[S7]} & How Mobile edge computing provides an IT service environment and cloud computing capabilities at the edge of the mobile network. & Edge computing application
        \\

        \hyperlink{[S8]}{[S8]} & How there is different parameters for implementation selection, and decision management for implementation of Decision Management. & Edge management
        \\

        \hyperlink{[S9]}{[S9]} & How edge computing relies on virtualization technologies and the measures to save possible threats against virtualization technologies. & Edge management
        \\

        \hyperlink{[S10]}{[S10]} & How edge computing is different in comparison to different  Computing techniques and how edge computing enhances different parameters. & Edge computing application
        \\

        \hyperlink{[S11]}{[S11]} & How edge computing is different from fog computing and the benfits for the same. & Edge computing application
        \\

        \hyperlink{[S12]}{[S12]} & How edge computing based CPS introduced a intelligent method to counter coupling problems and also helps to optimize sensors' utilization. & Edge Computing Security
        \\

        \hyperlink{[S13]}{[S13]} & How fog computing is more secure against different types of DDoS attacks, how fog computing provides not only provides network's edge but also security. & Edge Computing Security \\
        \\

        \hyperlink{[S14]}{[S14]} & Different types of IoT applications supported by mobile computing and there security requirements and vulnerabilities. & Edge computing application
        \\
         
        \hyperlink{[S15]}{[S15]} & Aims to present future research directions in security and technology development in a variety of Edge/Fog scenarios, with a focus on discussing the context's existing security challenges. & Edge Computing Security \\
        \\
        
        \hyperlink{[S16]}{[S16]} & How advancements in mobile edge computing have aided in the creation of innovative services are discussed, and WiCloud is suggested as a platform to support edge networking, proximity computing, and data collection for such services. & Edge Computing application
        \\
        
        \hyperlink{[S17]}{[S17]} & Two IoT applications that can take advantage of MEC's advantages are examined in this article for their security flaws. The first one uses an industrial IoT network as the foundation for an environment sensing technology. The second one, which is gaining momentum fast, is mobile IoT based on a network of unmanned aerial vehicles. & Edge computing for IoT
        \\
        
        \hyperlink{[S18]}{[S18]} & It provides an overview of the idea of edge computing and contrasts it with cloud computing. The edge computing architecture, keyword technology, security, and privacy protection, followed by a summary of the edge computing applications. & Edge computing security and application
        \\
        
        \hyperlink{[S19]}{[S19]} & focuses on the main categories of attacks in EC-assisted IoT and offers potential fixes and remedies along with the associated research initiatives. The next step is to categorise various security and privacy issues that have been raised in the literature using security services as well as security goals and functions. & Edge computing for IoT
        \\
        
        \hyperlink{[S20]}{[S20]} & In an edge computing environment where nodes communicate using the Message Queue Telemetry Transport (MQTT) protocol, the aim of this work is to investigate a distributed technique for mitigating Denial of Service (DoS) assaults on the fog node. & Edge computing security
        \\
        
        \hyperlink{[S21]}{[S21]} & In order to improve the security of the MEC system by identifying wireless network spoofing attacks, this study investigates the security threats in mobile edge computing (MEC) of the Internet of things. It then proposes a deep-learning (DL)-based physical (PHY) layer authentication scheme that uses CSI. & Edge computing security
        \\
        
        \hyperlink{[S22]}{[S22]} & seeks to establish the greatest possible match between user demands and available resources in order to achieve optimum resource allocation. The experimental results show that how the edge-based strategy is an effective means of addressing the coupling issue for cyber-physical systems. & Edge computing security and application
        \\
        
        \hyperlink{[S23]}{[S23]} & This article describes the many attack types that the Edge network faces in general, as well as the intrusion detection systems and accompanying machine learning techniques that address these security and privacy issues. & Edge computing security
        \\
        
        \hyperlink{[S24]}{[S24]} & While encouraging future conversations on the merger of edge intelligence and intelligent edge, i.e., Edge Deep Learning, this survey can assist readers in understanding the relationships between supporting technologies. & Edge computing security and application
        \\
        
        \hyperlink{[S25]}{[S25]} & Systematic reviews use a reliable, strict, and auditable technique in an effort to give an objective assessment of a research issue. & User Privacy and Edge Application
        \\
        
        \hyperlink{[S26]}{[S26]} & The replication demonstrated the value of the snowballing approach and demonstrates how the final result matched that of the first systematic literature studies. & Reference of SLR
        \\
        
        \hyperlink{[S27]}{[S27]} & How a new hybrid learning security framework uses machine learning power and the knowledge of security specialists to defend the edge computing network from known and undiscovered threats while reducing false detection rates & Blockchain SLR
        \\
        
        \hyperlink{[S28]}{[S28]} & a description of the IIoT's intelligent computing architecture including cloud computing and cooperative edge computing. & User Privacy and Edge Application \\
        
        \hyperlink{[S29]}{[S29]} & how to identify a malicious flow by looking at the fewest number of packets feasible. They analyse the model using both open data sets from prior research and data sets gathered from other sources in order to validate the suggested mechanism. & Edge computing security
        \\
        
        \hyperlink{[S30]}{[S30]} & It focuses on a Man-in-the-Middle (MITM) attack lab experiment that demonstrates how encrypted network data is analysed in-depth using supervised machine learning to identify social media apps. & Edge computing security
        \\
        
        \hyperlink{[S31]}{[S31]} & article examines the security threats associated with big data platforms used in the healthcare industry and how machine learning might help to reduce such risks. & Edge computing security
        \\
        
        \hyperlink{[S32]}{[S32]} &  It will provide a summary of the uses of deep learning at the network edge and talk about several methods for fast running deep learning inference on a combination of end devices, edge servers, and the cloud.  & Edge computing security
        \\
        
        \hyperlink{[S33]}{[S33]} & A smart home system simulation has been created for this investigation. The entire procedure would create a clear classifier to identify the distinction between regular and mutation codes, which could be used to detect network mutation codes. & Edge computing security
        \\
        
        \hyperlink{[S34]}{[S34]} & This demonstrates how the Internet of Things (IoT), networks and machine visualisation, public-key cryptography, web applications, certification schemes, and the secure storing of Personally Identifiable Information all lend themselves nicely to novel blockchain applications. & Edge computing security
        \\
        
         \hyperlink{[S35]}{[S35]} & In an effort to explain what has become a prevalent question as companies of all sizes compete to find the optimal place for computing power, this article compares and contrasts fog and edge computing and provides some real-world examples.& Edge computing security
         \\
    
        \hyperlink{[S36]}{[S36]} & This study takes use of current developments in the field of large-scale machine learning as well as the adaptability of cloud-based systems. & Edge computing security \\
\end{longtable}

\section{Future research direction of edge computing security}
\label{sec:FutureWork}
By referring the findings and observations of security of edge computing, we came up with the topics that will be worth digging into:

\textbf{ Edge computing efficiency:} We know that the task of edge computing it to provide seamless experience to the user when it comes to internet services \cite{a11,a12}. Mostly, edge servers are located more near to the user location to improve latency. Although we are using effective methods to transfer data, we can develop numerous ways to extract the only needed information from the user device. If numbers of devices are implemented using such method, we can have eventually smooth edge computing experience. Thus, more research and experiments can be performed to find out more efficient methods to extract the needed data. At the end, we can have more free bandwidth meaning less chances of data getting buffered. Plus, we can have better user privacy as only needed data are transferred for further processing. Thus, it is worth developing needed methods and algorithms for intended use.

\textbf{Edge Computing Security}: As we discussed, we know that many attacks are taking place when it comes to networking or edge computing. There are ways to fight against these attacks however, it is not so efficient as it will be not possible to detect attack using usual methods of doing. Machine learning bring some different approach to look upon a problem. Researchers have used the algorithms of machine learning on edge computing as well but we are still lacking of proper statistics to get the effect of machine learning on edge computing security. Hence, studies should be performed on deciding which method is suitable in what percentage. It helps us to decide which method to use in what kind of situation. Hence, we will be having optimal ways to use the optimal solution.

\section{Conclusion and Future Work}
In this SLR, we have seen what edge computing is, how it is used to improve cloud computing and to improve Quality of Service. Initially, the keyword provided in the title of this SLR is used for searching the related document and we found out the current problem faced by edge computing. This research thus listed out the common and usually seen problems/attack faced in edge computing. Also, the research listed the possible solution using latest technology such as machine learning. However, we have also seen that being a new technology requires more methods and results to come out so that it becomes trusted by everyone.

This section illustrates possible opportunities available for future research to carried out in the field of edge computing security. The increased demand of IoT devices bring drastic need of edge computing to offload the tasks of networking and computing. Thus, increased demand bring possibly more attacks on the system and hence the need to find more ways to make system secure as a whole.

\textbf{Potential research agenda 1:} The need of edge computing can be defined properly and need to work on algorithm and methods that can decrease the rate data is generating. This can be either done at user end or at cloud end. The methods to store and process user data locally are also needed to work on. User devices are becoming more powerful day by day, hence either performing real time process or calculating needed outcome locally on the device offloads a bigger portion from the cloud or edge. 

\textbf{Potential research agenda 2:} Several finding have identified different attacks caused on the edge servers and edge devices. The new approach of using machine learning to solve the problem seems more promising than the old method of solving problems. However, because of the new technology, there are not much of the resources available about ML to make the conclusion. By performing more number of research and experiments, we will be having more number of results and thus more statistics lead us to conclude better on which method to use for what kind of problem.  

\textbf{Potential research agenda 3:} The use of machine learning is not limited to just finding the attacks, it can be even used to extract possible pattern and methods of attack from the data collected till now. Hence, by using sub-branch of machine learning such as unsupervised machine learning, we can have more data exploration which can generate some useful information on possible reasons of attacks. This information can be used to design the architecture of the edge computing to make it more strong against the possible attacks caused on edge computing security.

\section*{Declaration of interest}
None.

\clearpage
\bibliographystyle{ieeetr}
\bibliography{references}

\begin{thebibliography}{10}

\bibitem{wu2019edge}
W.~Wu, Q.~Zhang, and H.~J. Wang, ``Edge computing security protection from the
  perspective of classified protection of cybersecurity,'' in {\em 2019 6th
  International Conference on Information Science and Control Engineering
  (ICISCE)}, pp.~278--281, IEEE, 2019.

\bibitem{yazdinejad2020energy}
A.~Yazdinejad, R.~M. Parizi, A.~Dehghantanha, Q.~Zhang, and K.-K.~R. Choo, ``An
  energy-efficient sdn controller architecture for iot networks with
  blockchain-based security,'' {\em IEEE Transactions on Services Computing},
  vol.~13, no.~4, pp.~625--638, 2020.

\bibitem{shi2016promise}
W.~Shi and S.~Dustdar, ``The promise of edge computing,'' {\em Computer},
  vol.~49, no.~5, pp.~78--81, 2016.

\bibitem{xiao2019edge}
Y.~Xiao, Y.~Jia, C.~Liu, X.~Cheng, J.~Yu, and W.~Lv, ``Edge computing security:
  State of the art and challenges,'' {\em Proceedings of the IEEE}, vol.~107,
  no.~8, pp.~1608--1631, 2019.

\bibitem{sha2020survey}
K.~Sha, T.~A. Yang, W.~Wei, and S.~Davari, ``A survey of edge computing-based
  designs for iot security,'' {\em Digital Communications and Networks},
  vol.~6, no.~2, pp.~195--202, 2020.

\bibitem{a2}
A.~Yazdinejad, A.~Dehghantanha, R.~M. Parizi, M.~Hammoudeh, H.~Karimipour, and
  G.~Srivastava, ``Block hunter: Federated learning for cyber threat hunting in
  blockchain-based iiot networks,'' {\em IEEE Transactions on Industrial
  Informatics}, 2022.

\bibitem{a3}
A.~Yazdinejad, A.~Dehghantanha, R.~M. Parizi, G.~Srivastava, and H.~Karimipour,
  ``Secure intelligent fuzzy blockchain framework: Effective threat detection
  in iot networks,'' {\em Computers in Industry}, vol.~144, p.~103801, 2023.

\bibitem{abawajy2018identifying}
J.~Abawajy, S.~Huda, S.~Sharmeen, M.~M. Hassan, and A.~Almogren, ``Identifying
  cyber threats to mobile-iot applications in edge computing paradigm,'' {\em
  Future Generation Computer Systems}, vol.~89, pp.~525--538, 2018.

\bibitem{myneni2022smartdefense}
S.~Myneni, A.~Chowdhary, D.~Huang, and A.~Alshamrani, ``Smartdefense: A
  distributed deep defense against ddos attacks with edge computing,'' {\em
  Computer Networks}, vol.~209, p.~108874, 2022.

\bibitem{ren2018edge}
J.~Ren, Y.~Pan, A.~Goscinski, and R.~A. Beyah, ``Edge computing for the
  internet of things,'' {\em IEEE Network}, vol.~32, no.~1, pp.~6--7, 2018.

\bibitem{ranaweera2021survey}
P.~Ranaweera, A.~D. Jurcut, and M.~Liyanage, ``Survey on multi-access edge
  computing security and privacy,'' {\em IEEE Communications Surveys \&
  Tutorials}, vol.~23, no.~2, pp.~1078--1124, 2021.

\bibitem{a5}
A.~Yazdinejad, M.~Kazemi, R.~M. Parizi, A.~Dehghantanha, and H.~Karimipour,
  ``An ensemble deep learning model for cyber threat hunting in industrial
  internet of things,'' {\em Digital Communications and Networks}, 2022.

\bibitem{satyanarayanan2017emergence}
M.~Satyanarayanan, ``The emergence of edge computing,'' {\em Computer},
  vol.~50, no.~1, pp.~30--39, 2017.

\bibitem{abbas2017mobile}
N.~Abbas, Y.~Zhang, A.~Taherkordi, and T.~Skeie, ``Mobile edge computing: A
  survey,'' {\em IEEE Internet of Things Journal}, vol.~5, no.~1, pp.~450--465,
  2017.

\bibitem{a4}
A.~Yazdinejad, B.~Zolfaghari, A.~Dehghantanha, H.~Karimipour, G.~Srivastava,
  and R.~M. Parizi, ``Accurate threat hunting in industrial internet of things
  edge devices,'' {\em Digital Communications and Networks}, 2022.

\bibitem{dolui2017comparison}
K.~Dolui and S.~K. Datta, {\em Comparison of edge computing implementations:
  Fog computing, cloudlet and mobile edge computing}.
\newblock IEEE, 2017.

\bibitem{yousefpour2019all}
A.~Yousefpour, C.~Fung, T.~Nguyen, K.~Kadiyala, F.~Jalali, A.~Niakanlahiji,
  J.~Kong, and J.~P. Jue, ``All one needs to know about fog computing and
  related edge computing paradigms: A complete survey,'' {\em Journal of
  Systems Architecture}, vol.~98, pp.~289--330, 2019.

\bibitem{paharia2020comprehensive}
B.~Paharia and K.~Bhushan, ``A comprehensive review of distributed denial of
  service (ddos) attacks in fog computing environment,'' {\em Handbook of
  computer networks and cyber security}, pp.~493--524, 2020.

\bibitem{shirazi2017extended}
S.~N. Shirazi, A.~Gouglidis, A.~Farshad, and D.~Hutchison, ``The extended
  cloud: Review and analysis of mobile edge computing and fog from a security
  and resilience perspective,'' {\em IEEE Journal on Selected Areas in
  Communications}, vol.~35, no.~11, pp.~2586--2595, 2017.

\bibitem{caprolu2019edge}
M.~Caprolu, R.~Di~Pietro, F.~Lombardi, and S.~Raponi, ``Edge computing
  perspectives: architectures, technologies, and open security issues,'' in
  {\em 2019 IEEE International Conference on Edge Computing (EDGE)},
  pp.~116--123, IEEE, 2019.

\bibitem{li2016mobile}
H.~Li, G.~Shou, Y.~Hu, and Z.~Guo, ``Mobile edge computing: Progress and
  challenges,'' in {\em 2016 4th IEEE international conference on mobile cloud
  computing, services, and engineering (MobileCloud)}, pp.~83--84, IEEE, 2016.

\bibitem{he2018security}
D.~He, S.~Chan, and M.~Guizani, ``Security in the internet of things supported
  by mobile edge computing,'' {\em IEEE Communications Magazine}, vol.~56,
  no.~8, pp.~56--61, 2018.

\bibitem{a6}
A.~Yazdinejad, R.~M. Parizi, A.~Dehghantanha, H.~Karimipour, G.~Srivastava, and
  M.~Aledhari, ``Enabling drones in the internet of things with decentralized
  blockchain-based security,'' {\em IEEE Internet of Things Journal}, vol.~8,
  no.~8, pp.~6406--6415, 2020.

\bibitem{cao2020overview}
K.~Cao, Y.~Liu, G.~Meng, and Q.~Sun, ``An overview on edge computing
  research,'' {\em IEEE access}, vol.~8, pp.~85714--85728, 2020.

\bibitem{alwarafy2020survey}
A.~Alwarafy, K.~A. Al-Thelaya, M.~Abdallah, J.~Schneider, and M.~Hamdi, ``A
  survey on security and privacy issues in edge-computing-assisted internet of
  things,'' {\em IEEE Internet of Things Journal}, vol.~8, no.~6,
  pp.~4004--4022, 2020.

\bibitem{potrino2019distributed}
G.~Potrino, F.~De~Rango, and P.~Fazio, ``A distributed mitigation strategy
  against dos attacks in edge computing,'' in {\em 2019 Wireless
  Telecommunications Symposium (WTS)}, pp.~1--7, IEEE, 2019.

\bibitem{liao2019security}
R.-F. Liao, H.~Wen, J.~Wu, F.~Pan, A.~Xu, H.~Song, F.~Xie, Y.~Jiang, and
  M.~Cao, ``Security enhancement for mobile edge computing through physical
  layer authentication,'' {\em IEEE Access}, vol.~7, pp.~116390--116401, 2019.

\bibitem{wang2020intelligent}
T.~Wang, Y.~Liang, Y.~Yang, G.~Xu, H.~Peng, A.~Liu, and W.~Jia, ``An
  intelligent edge-computing-based method to counter coupling problems in
  cyber-physical systems,'' {\em IEEE Network}, vol.~34, no.~3, pp.~16--22,
  2020.

\bibitem{singh2021machine}
S.~Singh, R.~Sulthana, T.~Shewale, V.~Chamola, A.~Benslimane, and B.~Sikdar,
  ``Machine-learning-assisted security and privacy provisioning for edge
  computing: A survey,'' {\em IEEE Internet of Things Journal}, vol.~9, no.~1,
  pp.~236--260, 2021.

\bibitem{a7}
A.~Yazdinejad, R.~M. Parizi, A.~Dehghantanha, and K.-K.~R. Choo,
  ``Blockchain-enabled authentication handover with efficient privacy
  protection in sdn-based 5g networks,'' {\em IEEE Transactions on Network
  Science and Engineering}, vol.~8, no.~2, pp.~1120--1132, 2019.

\bibitem{wang2020convergence}
X.~Wang, Y.~Han, V.~C. Leung, D.~Niyato, X.~Yan, and X.~Chen, ``Convergence of
  edge computing and deep learning: A comprehensive survey,'' {\em IEEE
  Communications Surveys \& Tutorials}, vol.~22, no.~2, pp.~869--904, 2020.

\bibitem{kitchenham2007guidelines}
B.~Kitchenham and S.~Charters, {\em Guidelines for performing systematic
  literature reviews in software engineering}, 2007.

\bibitem{wohlin2014guidelines}
C.~Wohlin, ``Guidelines for snowballing in systematic literature studies and a
  replication in software engineering,'' in {\em Proceedings of the 18th
  international conference on evaluation and assessment in software
  engineering}, pp.~1--10, 2014.

\bibitem{sedjelmaci2021trusted}
H.~Sedjelmaci, S.-M. Senouci, N.~Ansari, and A.~Boualouache, ``A trusted hybrid
  learning approach to secure edge computing,'' {\em IEEE Consumer Electronics
  Magazine}, vol.~11, no.~3, pp.~30--37, 2021.

\bibitem{a8}
A.~Yazdinejad, R.~M. Parizi, A.~Dehghantanha, and K.-K.~R. Choo,
  ``P4-to-blockchain: A secure blockchain-enabled packet parser for software
  defined networking,'' {\em Computers \& Security}, vol.~88, p.~101629, 2020.

\bibitem{a9}
A.~Yazdinejad, A.~Bohlooli, and K.~Jamshidi, ``Performance improvement and
  hardware implementation of open flow switch using fpga,'' in {\em 2019 5th
  Conference on Knowledge Based Engineering and Innovation (KBEI)},
  pp.~515--520, IEEE, 2019.

\bibitem{sun2019ai}
W.~Sun, J.~Liu, and Y.~Yue, ``Ai-enhanced offloading in edge computing: When
  machine learning meets industrial iot,'' {\em IEEE Network}, vol.~33, no.~5,
  pp.~68--74, 2019.

\bibitem{hwang2019detecting}
R.-H. Hwang, M.-C. Peng, and C.-W. Huang, ``Detecting iot malicious traffic
  based on autoencoder and convolutional neural network,'' in {\em 2019 IEEE
  Globecom Workshops (GC Wkshps)}, pp.~1--6, IEEE, 2019.

\bibitem{a10}
A.~Yazdinejad, R.~M. Parizi, A.~Bohlooli, A.~Dehghantanha, and K.-K.~R. Choo,
  ``A high-performance framework for a network programmable packet processor
  using p4 and fpga,'' {\em Journal of Network and Computer Applications},
  vol.~156, p.~102564, 2020.

\bibitem{al2020man}
A.~Al-Hababi and S.~C. Tokgoz, ``Man-in-the-middle attacks to detect and
  identify services in encrypted network flows using machine learning,'' in
  {\em 2020 3rd International Conference on Advanced Communication Technologies
  and Networking (CommNet)}, pp.~1--5, IEEE, 2020.

\bibitem{unal2022machine}
D.~Unal, S.~Bennbaia, and F.~O. Catak, ``Machine learning for the security of
  healthcare systems based on internet of things and edge computing,'' in {\em
  Cybersecurity and Cognitive Science}, pp.~299--320, Elsevier, 2022.

\bibitem{chen2019deep}
J.~Chen and X.~Ran, ``Deep learning with edge computing: A review,'' {\em
  Proceedings of the IEEE}, vol.~107, no.~8, pp.~1655--1674, 2019.

\bibitem{hou2019use}
S.~Hou and X.~Huang, ``Use of machine learning in detecting network security of
  edge computing system,'' in {\em 2019 IEEE 4th International Conference on
  Big Data Analytics (ICBDA)}, pp.~252--256, IEEE, 2019.

\bibitem{a11}
A.~Yazdinejad, A.~Bohlooli, and K.~Jamshidi, ``Efficient design and hardware
  implementation of the openflow v1. 3 switch on the virtex-6 fpga ml605,''
  {\em The Journal of Supercomputing}, vol.~74, no.~3, pp.~1299--1320, 2018.

\bibitem{a12}
A.~Yazdinejad, G.~Srivastava, R.~M. Parizi, A.~Dehghantanha, K.-K.~R. Choo, and
  M.~Aledhari, ``Decentralized authentication of distributed patients in
  hospital networks using blockchain,'' {\em IEEE journal of biomedical and
  health informatics}, vol.~24, no.~8, pp.~2146--2156, 2020.

\end{thebibliography}

\section*{Primary Studies}
\   

\hypertarget{[S1]}{[S1]}  \href{https://ieeexplore.ieee.org/abstract/document/9107702}{Edge computing security protection from the perspective of classified protection of cybersecurity}

\hypertarget{[S2]}{[S2]}  \href{https://ieeexplore.ieee.org/abstract/document/8741060?casa_token=JU2BfJ9tvCQAAAAA:lLGt4TUol6YofLjcROKlvUM9xCTLD-Cv9PS2V38sHTpWGwMH-6lfu39Otwr6PpM0NLSVlWemQDk}{Xiao, Yinhao, et al. "Edge computing security: State of the art and challenges." Proceedings of the IEEE 107.8 (2019): 1608-1631.
}

\hypertarget{[S3]}{[S3]}  \href{https://www.sciencedirect.com/science/article/pii/S0167739X18300906?casa_token=4D62QfLRrZoAAAAA:RwFSVUF7_3quscHEA7b2HOQIgP8znDtM1O_1dEsCQKMO73Zztx4XvCMc_K8VLrxSXrYsiEckNmY}{Abawajy, Jemal, et al. "Identifying cyber threats to mobile-IoT applications in edge computing paradigm." Future Generation Computer Systems 89 (2018): 525-538.
}

\hypertarget{[S4]}{[S4]}  \href{https://www.sciencedirect.com/science/article/pii/S1389128622000792?casa_token=4MxzEulMzEoAAAAA:ZdfYNyzifzEOPE9T3dwEjM48cZAbvKjpmHu_DOXO7jgi0NMDLTnR-sQJGt4ZPVbkXBPQk1rf5D8}{Myneni, Sowmya, et al. "SmartDefense: A distributed deep defense against DDoS attacks with edge computing." Computer Networks 209 (2022): 108874.
}

\hypertarget{[S5]}{[S5]}  \href{https://ieeexplore.ieee.org/abstract/document/7469991?casa_token=lBNMR4_QBAIAAAAA:pz1hnDh0Pg43Tg16coa4aXjeR8OhS_0FM3R0w9lJmJRFPrbD6nsJDLqG7J50mbM2cKXmfcWzjnw}{Shi, Weisong, and Schahram Dustdar. "The promise of edge computing." Computer 49.5 (2016): 78-81.
}

\hypertarget{[S6]}{[S6]}  \href{https://ieeexplore.ieee.org/abstract/document/7807196?casa_token=qGWMOfdzgC8AAAAA:5TH2FCUl-qliCVIORJwvTvu9javMvmKQbsXJAwKtmvBbpb2xTOxaU9uv4bBUjiqT1sX4KzBB-wg}{Satyanarayanan, Mahadev. "The emergence of edge computing." Computer 50.1 (2017): 30-39.
}

\hypertarget{[S7]}{[S7]}  \href{https://ieeexplore.ieee.org/abstract/document/9364272?casa_token=f7SD19WyfIQAAAAA:MVFd59F7Yh-Pwffrutr8SaFuqIBqzJYD_yCtTG5QgJib6dw2AYRxhYQ-_hAIvNwmnPF3Jn7Oh8A}{Ranaweera, Pasika, Anca Delica Jurcut, and Madhushanka Liyanage paper on "Survey on multi-access edge computing security and privacy."}

\hypertarget{[S8]}{[S8]}  \href{https://ieeexplore.ieee.org/abstract/document/8016213?casa_token=iSdJ7PJyM2EAAAAA:1BraWO0XCJtQJRxBiTjx4CFG_HN8ZpewXrdJCHy3yfs8BGxKinXNd6B1fIxob8H0OiqpvIeTEz8}{Dolui, Koustabh, and Soumya Kanti Datta. "Comparison of edge computing implementations: Fog computing, cloudlet and mobile edge computing." 2017 Global Internet of Things Summit (GIoTS). IEEE, 2017.
}

\hypertarget{[S9]}{[S9]}  \href{https://ieeexplore.ieee.org/abstract/document/8812214?casa_token=AKQPaT6WSOAAAAAA:0ViQb4UTp_V2IT8enSVeLFNBodX0fv_292LWnzuJqCDKYNX5GS66pscL4oPbu1KjLqv58wBO-1I}{Caprolu, Maurantonio, et al. "Edge computing perspectives: architectures, technologies, and open security issues." 2019 IEEE International Conference on Edge Computing (EDGE). IEEE, 2019.
}

\hypertarget{[S10]}{[S10]}  \href{https://www.sciencedirect.com/science/article/pii/S1383762118306349}{Yousefpour, Ashkan, et al. "All one needs to know about fog computing and related edge computing paradigms: A complete survey." Journal of Systems Architecture 98 (2019): 289-330.
}

\hypertarget{[S11]}{[S11]}  \href{https://www.onlogic.com/company/io-hub/fog-computing-vs-edge-computing/}{Understanding Fog Computing vs Edge Computing, By Andrew Overheid
Categories: Tech Explained Published On: April 15th, 2022
}

\hypertarget{[S12]}{[S12]}  \href{https://ieeexplore.ieee.org/abstract/document/9105928?casa_token=AgBMmTvYbfsAAAAA:dSNNYJyoKXa7GAvfcGSTIwudis145cVZ3MSb-DFLeuSQMQt1BKZqpuJFv4uoklunrqKySh2RruQ}{Wang, Tian, et al. "An intelligent edge-computing-based method to counter coupling problems in cyber-physical systems." IEEE Network 34.3 (2020): 16-22.}

\hypertarget{[S13]}{[S13]}  \href{https://link.springer.com/chapter/10.1007/978-3-030-22277-2_20}{Paharia, Bhumika, and Kriti Bhushan. "A comprehensive review of distributed denial of service (DDoS) attacks in fog computing environment." Handbook of computer networks and cyber security (2020): 493-524.
}

\hypertarget{[S14]}{[S14]}  \href{https://ieeexplore.ieee.org/abstract/document/8436046?casa_token=VBlLXD0TPZQAAAAA:lZU1dj2w8Ira_D354l7jrigvOr0j_krGXZIJL8mZ8gfHV-919_H9JJ09DvbDFNcvKplexIVVXj4}{He, Daojing, Sammy Chan, and Mohsen Guizani. "Security in the internet of things supported by mobile edge computing." IEEE Communications Magazine 56.8 (2018): 56-61.}

\hypertarget{[S15]}{[S15]}  \href{https://ieeexplore.ieee.org/abstract/document/8715543?casa_token=dwLR5rPrhs4AAAAA:kdwjrnGMPODStd27QuyRSczqkiIdDfuCGvklfkd4TKJ_4BDciIyMY1GR-4eKxwFsD84JU90mzOk}{Potrino, Giuseppe, Floriano De Rango, and Peppino Fazio. "A distributed mitigation strategy against DoS attacks in edge computing." 2019 Wireless Telecommunications Symposium (WTS). IEEE, 2019.
}

\hypertarget{[S16]}{[S16]}  \href{https://ieeexplore.ieee.org/abstract/document/8793124}{Liao, Run-Fa, et al. "Security enhancement for mobile edge computing through physical layer authentication." IEEE Access 7 (2019): 116390-116401.}

\hypertarget{[S17]}{[S17]}  \href{https://ieeexplore.ieee.org/abstract/document/8030322?casa_token=13Dp3cyRUMwAAAAA:Knf7hnSdoIRFSS6xerZfOVjtmPaxk2EtIuAwDMWwAOf-Sk0ChhCiQ-zmITevKtHanWPntNy9WSo}{Abbas, Nasir, et al. "Mobile edge computing: A survey." IEEE Internet of Things Journal 5.1 (2017): 450-465.
}

\hypertarget{[S18]}{[S18]}  \href{https://www.sciencedirect.com/science/article/pii/S2352864818303018}{Sha, Kewei, et al. "A survey of edge computing-based designs for IoT security." Digital Communications and Networks 6.2 (2020): 195-202.
}

\hypertarget{[S19]}{[S19]}  \href{https://ieeexplore.ieee.org/abstract/document/9083958}{Cao, Keyan, et al. "An overview on edge computing research." IEEE access 8 (2020): 85714-85728.}

\hypertarget{[S20]}{[S20]}  \href{https://ieeexplore.ieee.org/abstract/document/9163078?casa_token=ZssKOfKuhc4AAAAA:3u9CNo3rnmVyqeVAOljAviKuYcKTi_GcMRBCJuMM8e4oHesM_gXFd3TqV0pcakuKF52zMO3gZqM}{Alwarafy, Abdulmalik, et al. "A survey on security and privacy issues in edge-computing-assisted internet of things." IEEE Internet of Things Journal 8.6 (2020): 4004-4022.
}

\hypertarget{[S21]}{[S21]}  \href{https://ieeexplore.ieee.org/abstract/document/8060526?casa_token=K5JzJvWgDtMAAAAA:Eba2KMcpAAyerQlKV_n5vrgx-1VpFn0N2OMB3fsBi1ubFX6kFBMQEGVnNzQL4pmQ3ZHvnPchq7M}{Shirazi, Syed Noorulhassan, et al. "The extended cloud: Review and analysis of mobile edge computing and fog from a security and resilience perspective." IEEE Journal on Selected Areas in Communications 35.11 (2017): 2586-2595.
}

\hypertarget{[S22]}{[S22]}  \href{https://ieeexplore.ieee.org/abstract/document/7474412?casa_token=bCW9e4R0RYcAAAAA:IhTOoJ7p8oU9sLPU9CPRd0Uc977L4oWWWsPKM440ReNnEWbmpBsGe46k_t0Zv9CawHNkL1PXplg}{Li, Hongxing, et al. "Mobile edge computing: Progress and challenges." 2016 4th IEEE international conference on mobile cloud computing, services, and engineering (MobileCloud). IEEE, 2016.
}

\hypertarget{[S23]}{[S23]}  \href{https://ieeexplore.ieee.org/abstract/document/8270624}{Ren, Ju, et al. "Edge computing for the internet of things." IEEE Network 32.1 (2018): 6-7.}

\hypertarget{[S24]}{[S24]}  \href{https://www.sciencedirect.com/science/article/pii/S2352864818301536}{Taylor, Paul J., et al. "A systematic literature review of blockchain cyber security." Digital Communications and Networks 6.2 (2020): 147-156.
}

\hypertarget{[S25]}{[S25]}  \href{https://dl.acm.org/doi/pdf/10.1145/2601248.2601268?casa_token=xtvSCX0UkzsAAAAA:qnIupdRpEFstLjjAmcNX5isPohITZ7HvkcQVdmaBEBfzw1Kld-Y99RmJQY_5sl-2N4cfFKwwhlYTJA}{Wohlin, Claes. "Guidelines for snowballing in systematic literature studies and a replication in software engineering." Proceedings of the 18th international conference on evaluation and assessment in software engineering. 2014.}

\hypertarget{[S26]}{[S26]}  \href{https://www.researchgate.net/profile/Barbara-Kitchenham/publication/302924724_Guidelines_for_performing_Systematic_Literature_Reviews_in_Software_Engineering/links/61712932766c4a211c03a6f7/Guidelines-for-performing-Systematic-Literature-Reviews-in-Software-Engineering.pdf}{Keele, Staffs. Guidelines for performing systematic literature reviews in software engineering. Vol. 5. Technical report, ver. 2.3 ebse technical report. ebse, 2007.
}

\hypertarget{[S27]}{[S27]}  \href{https://ieeexplore.ieee.org/abstract/document/9490350?casa_token=09EovF_BnnQAAAAA:smfXa-VKwpZ7d5kjb9AGNLs_zXNygyj7QkZ0YWIUjbKIlQBIwqOJ_NuQOfQzMxOo0PZKkeKIW_Q}{Singh, Shivani, et al. "Machine-Learning-Assisted Security and Privacy Provisioning for Edge Computing: A Survey." IEEE Internet of Things Journal 9.1 (2021): 236-260.
}

\hypertarget{[S28]}{[S28]}  \href{https://ieeexplore.ieee.org/abstract/document/8976180?casa_token=6-JLDEjh6zEAAAAA:JBhBW3HAUv5RbS7c1qtHz7L3XvveWMZEz0-ItcxG-cz6D9q-UKCAz8inCu_NRXv2zHWnRAGdB9Y}{Wang, Xiaofei, et al. "Convergence of edge computing and deep learning: A comprehensive survey." IEEE Communications Surveys.}

\hypertarget{[S29]}{[S29]}  \href{https://www.sciencedirect.com/science/article/pii/S0743731518302004?casa_token=R4-I3cDxbvgAAAAA:x7D7Ne1B4JwfamRHHla9iOfuKtxkWGj-J85wuGif3oO09TlIf0ryFn29j7UUF0mx1PJA2OuVFvc}{Kozik, Rafał, et al. "A scalable distributed machine learning approach for attack detection in edge computing environments." Journal of Parallel and Distributed Computing 119 (2018): 18-26.}

\hypertarget{[S30]}{[S30]}  \href{https://www.sciencedirect.com/science/article/pii/B9780323905701000073}{Unal, Devrim, Shada Bennbaia, and Ferhat Ozgur Catak. "Machine learning for the security of healthcare systems based on Internet of Things and edge computing." Cybersecurity and Cognitive Science. Academic Press, 2022. 299-320.}

\hypertarget{[S31]}{[S31]}  \href{https://ieeexplore.ieee.org/abstract/document/8763885?casa_token=SrIf5qpMTU8AAAAA:U57vlHRWLHeOEXMtdSqHJ_NU5ziqbwjetZCdDsTzJAiZSbI3cJypgpG1HfCcNOEqYZIAdK80M-Q}{Chen, Jiasi, and Xukan Ran. "Deep learning with edge computing: A review." Proceedings of the IEEE 107.8 (2019): 1655-1674.}

\hypertarget{[S32]}{[S32]}  \href{https://ieeexplore.ieee.org/abstract/document/8713237}{Hou, Size, and Xin Huang. "Use of machine learning in detecting network security of edge computing system." 2019 IEEE 4th International Conference on Big Data Analytics (ICBDA). IEEE, 2019.}

\hypertarget{[S33]}{[S33]}  \href{https://ieeexplore.ieee.org/abstract/document/8713237}{Hou, Size, and Xin Huang. "Use of machine learning in detecting network security of edge computing system." 2019 IEEE 4th International Conference on Big Data Analytics (ICBDA). IEEE, 2019.
}

\hypertarget{[S34]}{[S34]}  \href{https://ieeexplore.ieee.org/abstract/document/8863729?casa_token=9SV4eIDS-xcAAAAA:K9G_eQpPX_u3JdHfsmrdOKIaFdzN9KCMc0VqojzXDh5W3DTL_ajVnczK9nacduV8UPgm4tSH8BU}{Sun, Wen, Jiajia Liu, and Yanlin Yue. "AI-enhanced offloading in edge computing: When machine learning meets industrial IoT." IEEE Network 33.5 (2019): 68-74.
}

\hypertarget{[S35]}{[S35]}  \href{https://ieeexplore.ieee.org/abstract/document/9024425}{Hwang, Ren-Hung, Min-Chun Peng, and Chien-Wei Huang. "Detecting IoT malicious traffic based on autoencoder and convolutional neural network." 2019 IEEE Globecom Workshops
}

\hypertarget{[S36]}{[S36]}  \href{https://ieeexplore.ieee.org/abstract/document/9199617}{Al-Hababi, Abdulrahman, and Sezer C. Tokgoz. "Man-in-the-middle attacks to detect and identify services in encrypted network flows using machine learning." 2020 3rd International Conference on Advanced Communication Technologies and Networking (CommNet). IEEE, 2020.
}

\end{document}